# Electromagnetic Characteristics of a Superconducting Magnet for 28GHz ECR Ion Source according to the Series Resistance of a Protection Circuit

**Hongseok Lee, Young Kyu Mo, Onyou Lee, Junil Kim, Seungmin Bang, Jong O Kang, Jonggi Hong[a], Seokho Nam[b], Sukjin Choi[c], In Seok Hong[c], Min Chul Ahn[d] and Hyoungku Kang**

*Department of Electrical Engineering, Korea National University of Transportation, Chungju-si 380-702, Korea*

[a]*Center for Analytical Instrumentation Development, Korea Basic Science Institute, Busan, Korea*

[b]*Applied Superconductivity Laboratory, Yonsei University, Seoul, Korea*

[c]*Rare Isotope Science Project, Institute for Basic Science, Daejon, 305-811, Korea*

[d]*Department of Electrical Engineering, Kunsan National University, Jeonbuk, Korea*

A linear accelerator, called RAON, has been being developed as a part of Rare Isotope Science Project (RISP) by Institute for Basic Science (IBS) [1]. The linear accelerator utilizes an electron cyclotron resonance (ECR) ion source for providing intense highly charged ion beams to the linear accelerator. 28GHz ECR ion source can extract heavy ion beams from proton to uranium. A superconducting magnet system for 28GHz ECR ion source is composed of hexapole coils and four solenoid coils made with low Tc superconducting wires of NbTi [2]. The electromagnetic force acts on the superconducting magnets due to the magnetic field and flowing current in case of not only normal state but also quench state [3]. In case of quench on hexapole coils, unbalanced flowing current among the hexapole coils is generated and it causes unbalanced electromagnetic force. Coil motions and coil strains in quench state are larger than those in normal state due to unbalanced electromagnetic force among hexapole coils. Therefore, analysis on electromagnetic characteristics of superconducting magnet for 28GHz ECR ion source according to the series resistance of protection circuit in case of quench should be conducted. In this paper, analysis on electromagnetic characteristics of superconducting hexapole coils for 28GHz ECR ion source according to the series resistance of protection circuit in case of quench is performed by using finite elements method (FEM) simulation.

Email: kang@ut.ac.kr

Fax: +82-43-841-5145

## I. INTRODUCTION

The RAON is the name of a linear accelerator which has been being developed as a part of RISP by IBS. An ECR ion source is used for providing intense highly charged ion beams to the linear accelerator. The RAON utilizes 28GHz ECR ion source to provide various heavy ion beams from proton to uranium. 28GHz ECR ion source should use a superconducting magnet to confine intense highly charged ion beam. A superconducting magnet system for 28GHz ECR ion source is composed of hexapole coils nested inside four solenoid coils made with low Tc superconducting wires of NbTi. The superconducting magnets provide an axial magnetic field from four solenoid coils and a radial magnetic field from hexapole coils to confine ECR plasma stream. The electromagnetic force acts on the superconducting magnets due to the magnetic field and flowing current in case of not only normal state but also quench state. In case of quench on hexapole coils, unbalanced electromagnetic force is caused from unbalanced flowing current among hexapole coils and it can cause large coil motions and coil strains. Therefore, analysis on electromagnetic characteristics of superconducting magnets for 28GHz ECR ion source in case of quench should be conducted. In this paper, analysis on electromagnetic characteristics of superconducting hexapole coils with respect to the series resistance of protection circuit is performed by using FEM simulation in case of quench. It is expected that the results can be helpful to perform the design of superconducting magnet system for 28GHz ECR ion source for the RAON.

**II. THE SUPERCONDUCTING MAGNET SYSTEM FOR 28GHz ECR ION SOURCE**

The superconducting magnet system for 28GHz ECR ion source is composed of hexapole coils and four solenoid coils. For the 28GHz ECR ion source, the requirement of the axial magnetic field at the injection side ($B_{inj}$), extraction side ($B_{ext}$), and radial magnetic field at the plasma chamber surface ($B_r$) are 3.5, 2.0~2.5, and 2T, respectively [3].

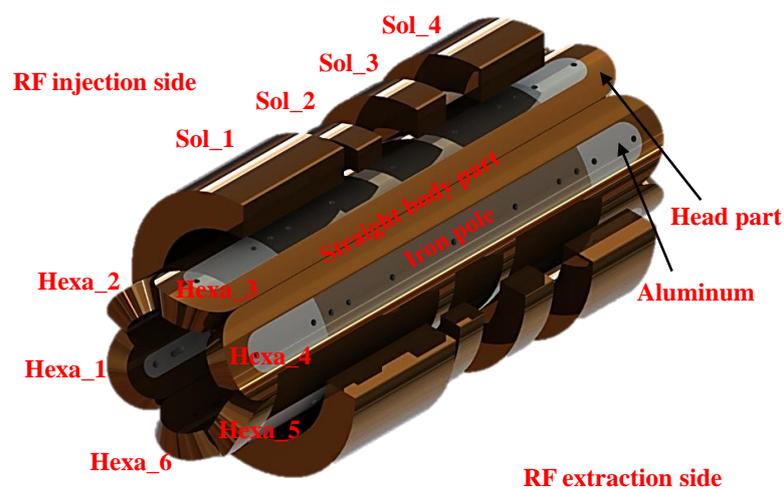

Fig. 1. Schematic view of the superconducting magnet system for 28GHz ECR ion source.

A conceptual schematic view of the superconducting magnet system is shown in Fig. 1 and the design parameters are summarized in Table 1. The large electromagnetic force on the end of hexapole coils is caused due to flowing current at the return sections and magnetic field. Especially, the electromagnetic force at the injection side is larger than that at extraction side [3]. In order to alleviate the electromagnetic force on the end of the hexapole coils at the injection side, the hexapole coils are moved 20mm from the center point to the injection side. An iron pole made of ferromagnetic iron is inserted inside each hexapole coils. The iron pole enhances radial magnetic field generated from the hexapole coils [2]. Fig. 2 shows the axial and radial magnetic field distribution of the definitive superconducting magnet system.

Table 1. Design parameters of a superconducting magnet system

| Solenoid | Sol_1 | Sol_2 | Sol_3 | Sol_4 | Hexapole coils |
|---|---|---|---|---|---|
| Axial position of center (mm) | -250 | -76 | 65 | 250 | |
| Inner radius (mm) | 188 | 188 | 188 | 188 | 108 |
| Depth (mm) | 67 | 45 | 58 | 67 | 50 |
| Width (mm) | 230 | 55 | 65 | 145 | |
| Superconductor | NbTi | NbTi | NbTi | NbTi | NbTi |
| Conductor size (mm) | 1.6*0.91 | 1.6*0.91 | 1.6*0.91 | 1.6*0.91 | 1.43*0.98 |
| Cu/NbTi ratio | 3.65 | 3.65 | 3.65 | 3.65 | 3 |
| Turns/coil | 8945 | 1436 | 2188 | 5639 | 1367 |
| Ic at 7 T (A) | 404 | 404 | 404 | 404 | 540 |
| Design Current (A) | -165 | 125 | 140 | -144 | 254 |
| $B_{max}$ (T) at coil at operating current | 5.73 | 4.72 | 3.65 | 4.65 | 6.79 |
| Wire length (km) | 12.46 | 1.9 | 2.98 | 7.85 | 2.56km/ 1 unit |

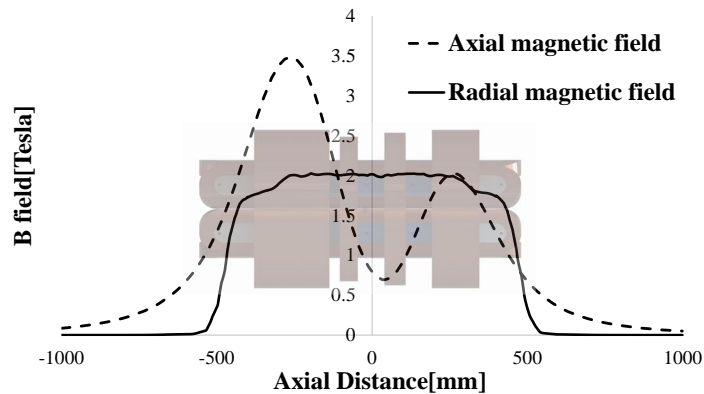

Fig. 2. Axial and radial magnetic field distributions of definitive superconducting magnet along the beam axis.

## III. CURRENT DISTRIBUTIONS WITH RESPECT TO SRIES RESISTANCE OF PROTECTION CIRCUIT IN QUENCH STATE

In case of quench on hexapole coils, unbalanced flowing current among hexapole coils is generated and it causes unbalanced electromagnetic force. Transient characteristics of hexapole coils are analyzed to estimate unbalanced flowing current. In order to analyze the transient characteristics of hexapole coils in quench state, a numerical analysis code is developed by using MATLAB [4].

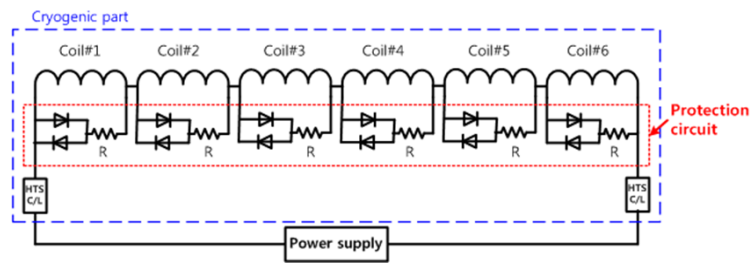

Fig. 3. Protection circuit of the hexapole coils.

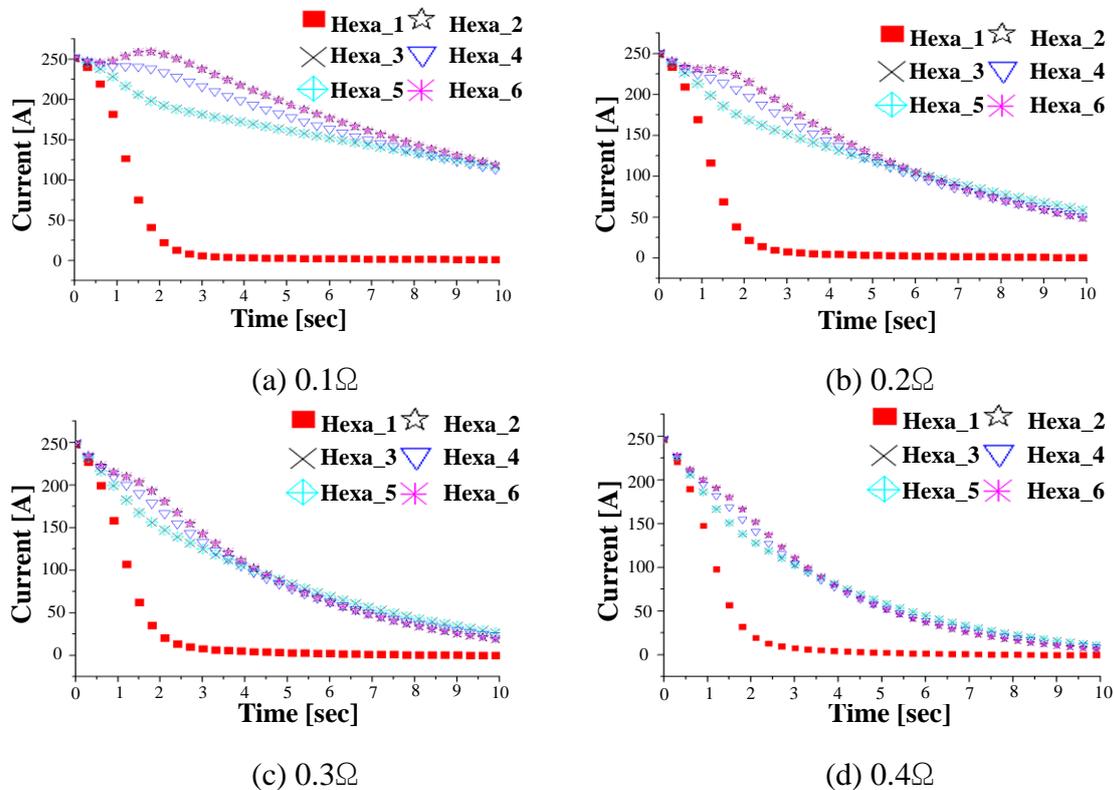

(a) 0.1Ω  (b) 0.2Ω
(c) 0.3Ω  (d) 0.4Ω

Fig. 4. Current distributions of the hexapole coils according to the series resistance of protection circuit in quench state.

Fig. 3 shows protection circuit of hexapole coils and Fig. 4 shows the results of the flowing current on hexapole coils over elapsed time according to the series resistance of protection circuit in case of quench. Fig. 4 (a), (b), (c), and (d) are 0.1, 0.2, 0.3, and 0.4Ω of the series resistance, respectively. The quench only occurs on a hexapole coil_1 (Hexa_1). As shown in Fig. 4, the flowing current decreases in quench state as the elapsed time increases. The difference of unbalanced flowing current among hexapole coils is largest at 2sec and prominently decrease as the series resistance of protection circuit increases.

## Ⅳ. ANALYSIS ON ELECTROMAGNETIC FORCE OF SUPERCONDUCTING MAGNET

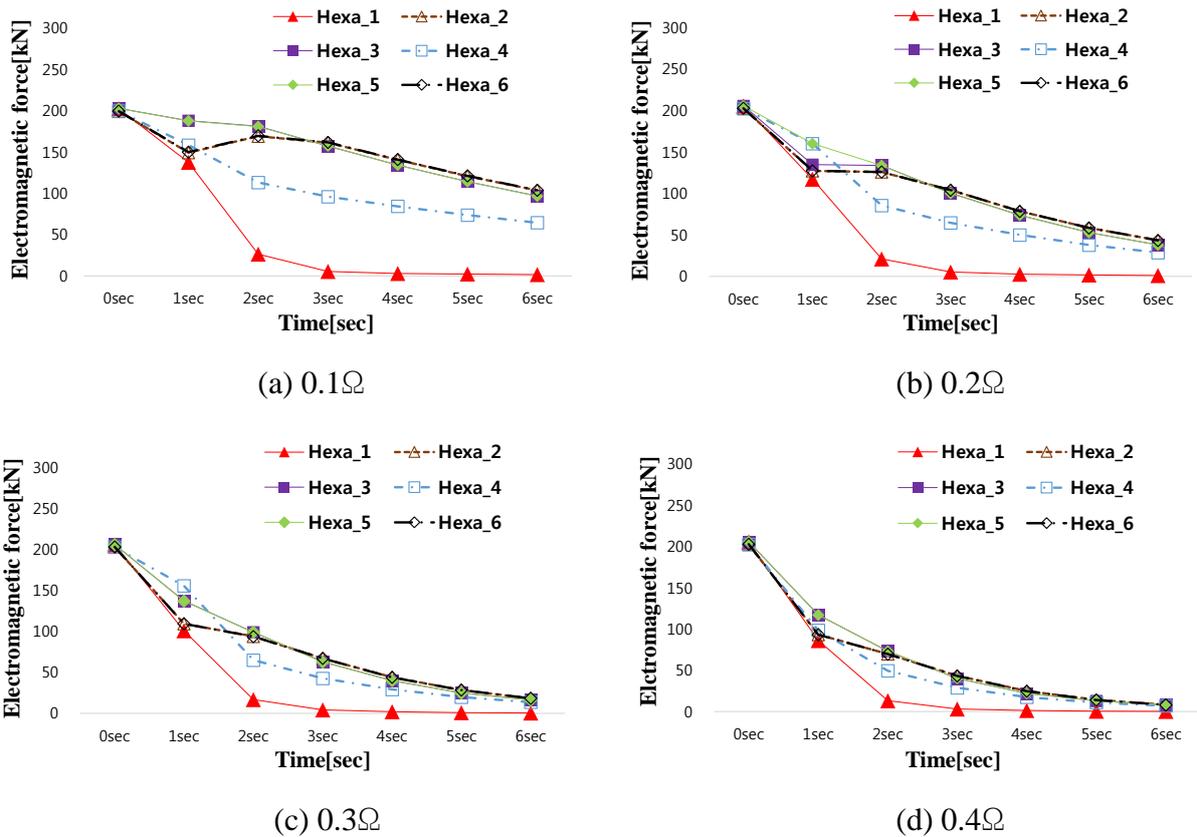

(a) 0.1Ω  (b) 0.2Ω  (c) 0.3Ω  (d) 0.4Ω

Fig. 5. Electromagnetic force on the hexapole coils according to the series resistance of protection circuit in quench state.

The electromagnetic force on hexapole coils according to the series resistance of protection circuit is calculated in case of quench. Fig. 5 shows the magnitude of the electromagnetic force on the hexapole coils according to the series resistance of protection circuit in case of quench on Hexa_1. As shown in Fig. 5, the electromagnetic force decreases in quench state as the elapsed time increases. The difference of unbalanced electromagnetic force among hexapole coils is largest at 2sec and prominently decrease as the series resistance of protection circuit increases. In order to deduce appropriate the series resistance of protection circuit, the electromagnetic force on hexapole coils according to the series resistance is compared. Fig. 6 shows comparison of the electromagnetic force according to the series resistance of protection circuit from 1sec to 4sec, in the quench state.

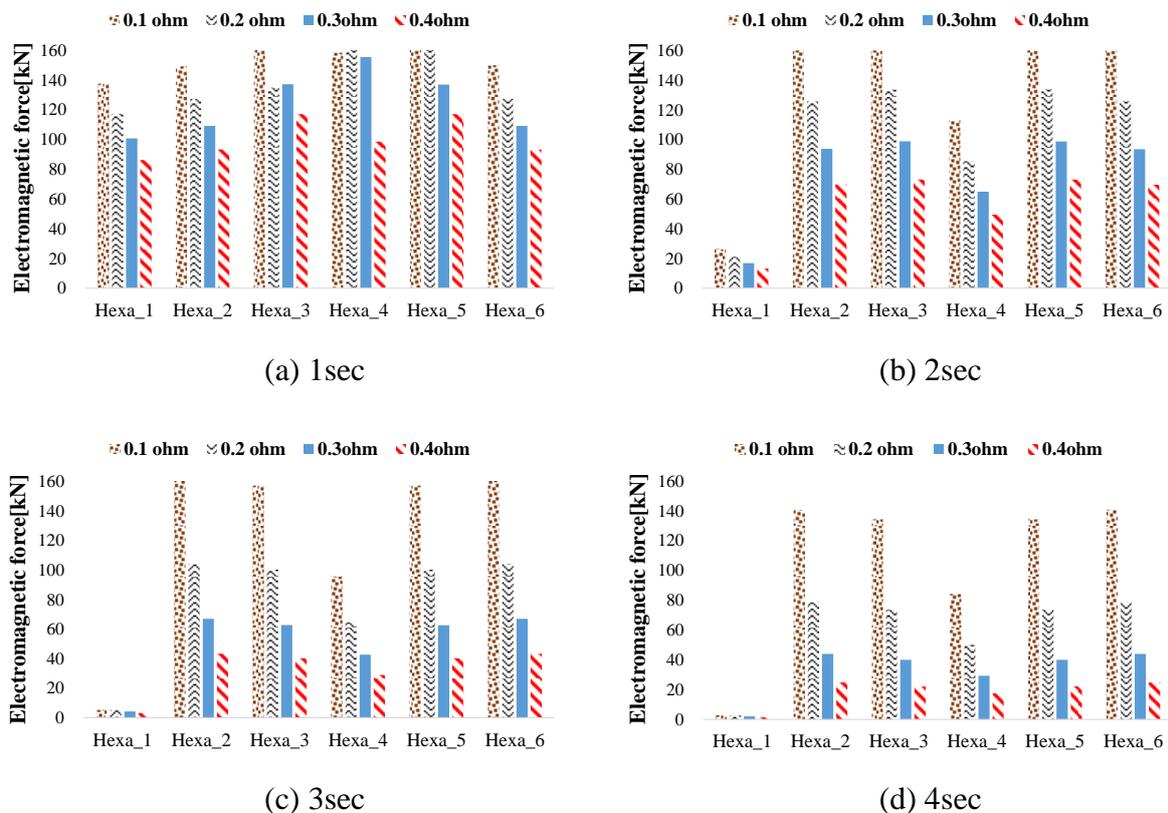

(a) 1sec    (b) 2sec

(c) 3sec    (d) 4sec

Fig. 6. Comparison of the electromagnetic force according to the series resitance of protection circuit from 1sec to 4sec in the quench state.

As shown in Fig. 6, 0.4Ω of the series resistance shows the lowest results of the electromagnetic force. As results, the magnitude of electromagnetic force on hexapole coils decreases as the series resistance of protection circuit increases.

## Ⅴ. CONCLUSIONS

In case of quench, the study on the electromagnetic characteristics of the superconducting magnets for 28GHz ECR ion source is performed by using FEM simulation. In order to analyze the transient characteristics of hexapole coils in case of quench, a numerical analysis code is developed by using MATLAB. Analysis on the transient characteristics of the hexapole coils according to the series resistance of protection circuit are conducted in quench state and the electromagnetic characteristics of the hexapole coils are analyzed. As results, it is found that the flowing current of the hexapole coils in quench state decreases as the elapsed time increases. It is found that the difference of unbalanced flowing current among the hexapole coils is largest at 2sec and prominently decrease as the series resistance of protection circuit increases. It is found that the electromagnetic characteristics among the hexapole coils conform to the flowing current characteristics. Also, it is found that 0.4Ω of the series resistance shows the lowest results of the unbalanced electromagnetic force. However, 0.2Ω of the series resistance considering voltage on superconducting magnets is determined, because the magnitude of voltage on superconducting magnets is below 500V. The research on the stress, deformation, and supporting structure will be dealt in another successive paper.

## ACKNOWLEDGEMENT

This work was supported by the Rare Isotope Science Project of Institute for Basic Science funded by Ministry of Science, ICT and Future Planning and National Research Foundation of Korea (2011-0032011)